# Calculation of Energy Spectrum of $^{12}$C Isotope by Relativistic Cluster model

Nafiseh Roshanbakht[1], Mohammad Reza Shojaei[1]
1. Department of physics, Shahrood University of Technology
P.O. Box 36155-316, Shahrood, Iran
E-mail: nroshanbakht@yahoo.com

**Abstract.** In this paper, we have calculated the energy spectrum of $^{12}$C isotope by cluster model. The experimental results show that the "Hoyle" state at 7.65 MeV in $^{12}$C isotope has a well-developed three-alpha structure. Hence, we select a three-body system and for interaction between the clusters we use modified Yukawa potential plus coulomb potential. Then, we solve the relativistic Klein-Gordon equation using Nikiforov-Uvarov method to calculate the energy spectrum. Finally, the calculated results are compared with the experimental data. The results show that the isotope $^{12}$C should be considered as consisting of three-alpha cluster and the modified Yukawa potential is adaptable for cluster interactions.

## 1. Introduction

The energy levels of a nucleus provide an important experimental evidence of the nuclear structure. Therefore, all the nuclear models, which have evolved from 1930s to now, are expected to reproduce the energy levels. It started from the uniform model of Wigner [1], the liquid drop model of Bohr [2], and the shell model [3]. The shell model is associated with the quanta properties of the nucleus and predicts some of them, such as spin and parity of nuclei states. The nuclear shell model is able to give sequence magic numbers, but for the nucleus which is away from closed shell, it cannot determine nuclear properties well. To alleviate these problems in the shell model, the independent particle model and the cluster model have been proposed. The cluster model has a long history too [4]. The cluster interpretation is a suitable model to describe nuclear states until now and has been successful in reproducing the energy spectra and other nuclear properties such as electro-magnetic properties, α-emission widths and α-particle elastic scattering data in nuclei near the double closed shell. In 1936 when Bethe predicted that nuclei are made of alpha particles and also proposed a geometrical arrangement of alpha particles inside nuclei, the cluster model was introduced [5]. In 1937, Wheeler [5] extended this work, and similar models suggested concurrently by Wefelmeier [7], Weizsacker [8], and Fano [9]. Freer and Merchant in 1997, studied the role of clustering and cluster models in nuclear reactions and examined the evidence for α-cluster chain configurations in the light even - even nuclei from $^8$Be to $^{28}$Si [10]. The alpha particle, $^4$He nucleus, is the most common cluster which is exceptionally stable (its first excited state is 20.2 MeV [11]), and its binding energy per nucleon is 7.07 MeV/nucleon. It is also very compact (a charge radius of 1.673fm [12]), and has zero spin for all quantum numbers (spin and isospin), which makes it easy to combine into larger systems. Also, it is the first doubly magic nucleus with the first closed shell $1S_{1/2}$, which accounts for its exceptional stability.

Recently, several systems of even-even nucleus were studied with a cluster model and their results were reasonably compared to the experimental spectra for the ground state and some excited states [13-16].

The core of this work is at least inspired by cluster models and is organized as follow. In section 2, we review the parametric generalization Nikiforov-Uvarov (NU) method briefly. We introduced the Klein-Gordon equation and solved it by the modified Yukawa potential plus coulomb to calculate spectra for the ground state and some excited states of $^{12}$C isotope in section 3. At the end, conclusion is given in section 4.

## 2. Nikiforov-Uvarov Method

The solution of the Klein-Gordon equation (KGE) similar to Schrödinger equation (SE) is a very important issue to solve many problems of nuclear and high-energy physics especially relativistic problems and to obtain an accurate solution of them is only possible for a few potentials. The standard analytical method to solve Schrödinger equation, Dirac equation and the Klein–Gordon equation with a variable coefficient is to expand the solution in a power series of the independent variable **r** and then find the recursion relationships for all the expansion coefficients [17]. This method has more details to reach the solution. Numerical and analytical methods complement each other to find an exact or approximate solution of the quantum and each would be much poorer without the other. But in our article, we use simple "hand-power methods" namely analytical methods because it more revealing to see the solution stages of the problem and so it would be more meaningful than the numerical solution.

The NU method is a powerful mathematical tool to solve Schrödinger, Dirac, Klein-Gordon wave equations for certain kinds of potentials. This method is based on the solutions of general second order linear differential equation with special orthogonal functions that starts by considering the following differential equation [18-20].

$$\Psi''(s) + \frac{\tilde{\tau}(s)}{\sigma(s)}\Psi'_n(s) + \frac{\tilde{\sigma}(s)}{\sigma^2(s)}\Psi_n(s) = 0 \tag{4}$$

where σ(s) and $\tilde{\sigma}(s)$ are polynomials (at most second-degree), and $\tilde{\tau}(s)$ is a first-degree polynomial. The following equation is a general form of the Schrödinger-like equation written for any potential [19]:

$$\left[\frac{d^2}{ds^2} + \frac{\alpha_1 - \alpha_2 s}{s(1-\alpha_3 s)}\frac{d}{ds} + \frac{-\xi_1 s^2 + \xi_2 s - \xi_3}{[s(1-\alpha_3 s)]^2}\right]\Psi_n(s) = 0 \tag{5}$$

According to the Nikiforov-Uvarov method, the energy spectrum function and the eigenfunctions are obtained by the following equations, respectively:

$$\alpha_2 n - (2n+1)\alpha_5 + (2n+1)(\sqrt{\alpha_9} + \alpha_3\sqrt{\alpha_8}) + n(n-1)\alpha_3 + \alpha_7 + 2\alpha_3\alpha_8 + 2\sqrt{\alpha_8\alpha_9} = 0 \tag{6}$$

and:

$$\Psi(s) = s^{\alpha_{12}}(1-\alpha_3 s)^{-\alpha_{12}-\frac{\alpha_{13}}{\alpha_3}} P_n^{(\alpha_{10}-1, \frac{\alpha_{11}}{\alpha_3}-\alpha_{10}-1)}(1-2\alpha_3 s) \tag{7}$$

Here, alpha functions are given by:

$$\alpha_4 = \frac{1}{2}(1-\alpha_1) \qquad \alpha_5 = \frac{1}{2}(\alpha_2 - 2\alpha_3) \qquad \alpha_6 = (\alpha_5^2 + \xi_1)$$

$$\alpha_8 = \alpha_4^2 + \xi_3 \qquad \alpha_9 = \alpha_3\alpha_7 + \alpha_3^2\alpha_8 + \alpha_6 \tag{8}$$

and:

$$\alpha_{10} = \alpha_1 + 2\alpha_4 + 2\sqrt{\alpha_8} \qquad \alpha_{11} = \alpha_2 - 2\alpha_5 + 2(\sqrt{\alpha_9} + \alpha_3\sqrt{\alpha_8})$$

$$\alpha_{12} = \alpha_4 + \sqrt{\alpha_8} \qquad \alpha_{13} = \alpha_5 - (\sqrt{\alpha_9} + \alpha_3\sqrt{\alpha_8}) \tag{9}$$

In some problems $\alpha_3 = 0$. For this type of problems when:

$$\lim_{\alpha_3 \to 0}(1-\alpha_3 s)P_n^{(\alpha_{10}-1, \frac{\alpha_{11}}{\alpha_3}-\alpha_{10}-1)} = L_n^{\alpha_{10}-1}(\alpha_{11} s) \tag{10}$$

and:

$$\lim_{\alpha_3 \to 0}(1-\alpha_3 s)^{-\alpha_{12}-\frac{\alpha_{13}}{\alpha_3}} = e^{\alpha_{13} s} \tag{11}$$

The solution given in Equation (4) becomes as [18-20]:

$$\Psi(s) = s^{\alpha_{12}} e^{\alpha_{13} s} L_n^{\alpha_{10}-1}(\alpha_{11} s) \tag{12}$$

## 3. Determination of energy spectrum of $^{12}$C isotope in the relativistic cluster model

The structure of the $^{12}$C nuclei has been one of the most important subjects in many studies, both experimental and theoretical, and also many successful results have been obtained in various cluster structures [21-26]. One typical cluster state of $^{12}$C is the famous Hoyle state $0_2^+$ (E=7.65 MeV). The state was predicted by Hoyle to account for the abundance of carbon in the universe and it was subsequently measured at an energy which was extremely close to that predicted [27]. Carbon is synthesized in stellar environments, first two alpha-particles admix momentarily to form $^8$Be and then immediately before the system decays a third α-particle is captured. This process occurs mainly through the 7.65 MeV, $0^+$, state [28]. This state is known to possess an extremely large radius, which can suppose the α-particles to retain their quasi-free characteristics. Here, we discuss $^{12}$C structure in three alpha clusters and the internal structure of the alphas and effects of the Pauli principle between the nucleons in the alpha clusters are negligible, but modeling the effective interaction among clusters is very important. The cluster-core interaction leads to the identification of clustering in the nuclear matter and the description of clustering phenomenon in various nuclei. In 1960s, Ali and Bodmer [4, 29] used the experimental data on alpha-alpha scattering and obtained potentials which were fitted to the scattering phase shifts. All their potentials had a repulsive part with strength $V_{0\ell}$ which was dependent on the angular momentum $\ell$ and an attractive part with a constant strength $V_\ell$ [4, 29], so that:

$$V(r) = V_{0l} \exp(-r^2/a^2) - V_l \exp(-r^2/a^2) \tag{13}$$

At a more microscopic level, the core-cluster interaction may be constructed from a nucleon-nucleon interaction. During the last decade, the modified phenomenological Saxon-Woods plus Cubic Saxon-Woods cluster potential has successfully described various phenomenon related to alpha clustering in light as well as even-even heavy nuclei. The form of the potential is given by:

$$V(r) = V_0 \left[ \frac{x}{1+\exp(\frac{r-R}{a})} + \frac{1-x}{1+\exp(\frac{r-R}{a})^3} \right] \tag{14}$$

This potential is parameterized in terms of the potential depth $V_0$, nuclear radius $R$, diffuseness $a$, and $x$ is a mixing parameter [30]. Prior to the development of the Saxon-Wood plus Saxon-Wood cubic potential form, such a microscopic interaction had been employed in various forms to describe cluster bound states in light nuclei [31] and the exotic decays in heavy nuclei [32]. Despite its success, modified Saxon-Wood potential model tells us very little about the microscopic nature of clustering in closed shell nuclei. The Yukawa interaction, in theoretical nuclear physics, was considered as the phenomenological central potential between nucleons and was widely used in modern physics and became an important physical model for theoretical studies [33-38]. So, it is an ideal potential which may be used in the cluster models.

In this paper, for interaction between the clusters, we select modified Yukawa potential plus coulomb repulsion as the central potential $V_c(r)$. Yukawa potential is short-range, so is an ideal potential for interaction between the clusters.

$$V_c(r) = -V_0 \frac{\exp(-\alpha r)}{r} + V_1 \frac{\exp(-\alpha r)}{r^2} + \frac{k}{r} \tag{15}$$

This potential has an attractive part with constant $V_0$, and a repulsive part with constant $V_1$ (because the nuclear forces saturate at very small distances), and k is the coulomb repulsion potential coefficient between the clusters.

As mentioned, the structure of "Hoyle" state, the first excited $0^+$ state, at -84.51 MeV in $^{12}$C isotope is influenced by clustering or the symmetries thereof. So, the system can be constructed from a variety of geometric arrangements of three-alpha particles. It might be expected that the compact equilateral-triangle arrangement is the lowest energy configuration [5, 39]. For our purposes, three identical body forces of the internal particle motion are described in terms of the Jacobi relative coordinate $\rho, \lambda$ and $R$ is center of mass. Now, we can introduce the hyper-radius quantity $x$ and the hyper-angle $\xi$ as follow [40-43]:

$$x = \sqrt{\rho^2 + \lambda^2} \qquad \xi = \tan\left(\frac{\rho}{\lambda}\right) \tag{16}$$

where:

$$\rho = \frac{\vec{r_1} - \vec{r_2}}{\sqrt{2}} \qquad \lambda = \frac{\vec{r_1} + \vec{r_2} - 2\vec{r_3}}{\sqrt{6}} \qquad R = \frac{\vec{r_1} + \vec{r_2} + \vec{r_3}}{\sqrt{3}} \tag{17}$$

and $\vec{r_1}$, $\vec{r_2}$ and $\vec{r_3}$ are relative positions three of the particle. Now, our central potential in new coordinate is:

$$V_c(x) = -V_0 \frac{\exp(-\alpha x)}{x} + V_1 \frac{\exp(-\alpha x)}{x^2} + \frac{k}{x} \tag{18}$$

where α, $V_0$, $V_1$ and k are constants. In terminology, the potentials S(x) and V(x) are respectively called vector and scalar potentials [44-47]. The reason is that the so-called scalar potential is bracketed with the mass and the so-called vector potential goes with the energy in Klein-Gordon equation.

The D-dimensional time-independent arbitrary l-states radial Klein-Gordon equation with scalar and vector potentials S(r) and V(r), where r = |r| is describing a spinless particle (such as α particle), takes the general form [44-47]:

$$\nabla_D^2 \psi_{l_1...l_{D-2}}^{(l_{D-1}=l)}(r) + \frac{1}{\hbar^2 c^2}\left\{[E_{nl} - V(r)]^2 - [Mc^2 + S(r)]^2\right\}\psi_{l_1...l_{D-2}}^{(l_{D-1}=l)}(r) = 0 \qquad (19)$$

where $E_{nl}$, M and $\nabla_D^2$ denote the KG energy, the mass and the D-dimensional Laplacian, respectively. In addition, **x** is a D-dimensional position vector in Jacobi coordinates. Therefore, by choosing a common ansatz for the wave function in the form of:

$$R_l(x) = x^{-(D-1)/2} u_l(x) \qquad (20)$$

Eq. (19) is reduced to Eq. (21), and KG equation turns into Schrödinger-like equation, and consequently the bound state solutions are simply obtained with NU method [48]:

$$\frac{d^2 u_l(x)}{dx^2} + \frac{1}{\hbar^2 c^2}\left\{[E_{nl} - V(x)]^2 - [Mc^2 + S(x)]^2 - \frac{(D+2l-1)(D+2l-3)}{4x^2}\right\}u_l(x) = 0 \qquad (21)$$

By replacing scalar and vector potentials S(r) and V(r) from Eq. (18) into Eq. (21), we have:

$$\frac{d^2 u_l(x)}{dx^2} + \frac{1}{\hbar^2 c^2}\left\{E_{nl}^2 - m^2 c^4 - 2(E_{nl} - mc^2)(e^{-\alpha x}(\frac{-V_0}{x} + \frac{V_1}{x^2}) + \frac{k}{x}) - \frac{(D+2l-1)(D+2l-3)}{4x^2}\right\}u_l(x) = 0 \qquad (22)$$

For giving to eigenvalues and corresponding eigenfunctions of the radial part, i.e., solution of Eq. (22), we rewrite it as follows:

$$\left[\frac{d^2}{dx^2} + \frac{5}{x}\frac{d}{dx} + \frac{-\xi_1 x^2 + \xi_2 x - \xi_3}{x^2}\right]\Psi_n(x) = 0 \qquad (23)$$

where:

$$\xi_1 = \frac{m^2 c^4 - E_{nl}^2 + 2(mc^2 + E_{nl})\alpha V_0}{\hbar^2 c^2} \qquad (24)$$

$$\xi_2 = -\frac{2(mc^2 + E_{nl})(-V_0 - \alpha V_1 + k)}{\hbar^2 c^2} \qquad (25)$$

$$\xi_3 = \frac{2V_1(mc^2 + E_{nl}) + l(l+4)\hbar^2 c^2}{\hbar^2 c^2} \qquad (26)$$

Now, we follow the Eq. (23) in the NU method after using Eq. (6). As a result, the exact energy eigenvalues function is obtained as:

$$(2n+1) - \frac{\xi_2}{\sqrt{\xi_1}} + 2\sqrt{4+\xi_3} = 0 \qquad (27)$$

In this model, the first excited $_1 0^+$ state and the first excited $_1 3^-$ state are fitted to the experimental data. The chosen parameters are $V_0$ = 50.14 MeV, $V_1$ = 6.13 MeV, k = 5.75 MeV and $\alpha$ = 0.12 fm$^{-1}$. The binding energy with this configuration obtained -92.22 MeV near compared to the experimental value. The measured binding energy in $^{12}$C isotope is equal to -92.16 MeV [49]. For the first and second excited states $2^+$ and $0^+$, the measured results are -87.72 MeV and -84.51 MeV and we obtain -87.78 MeV and -84.57 MeV, respectively. The difference between them is too small. Our model completely describes the next excited states $3^-$, $1^-$ and $4^+$, but in the second excited states $_2 2^+$ and $_2 0^+$ have not been fitted to the experimental data very well. The energy level $_2 2^+$ is -79.72 MeV and we obtain -78.48 MeV, also, the energy level $_2 0^+$ is -74.40 MeV and we obtain -76.00 MeV. Furthermore, the results in figure 1 and compared with the experimental data [50] shows that the $^{12}$C isotope should be considered as consisting of three-alpha cluster.

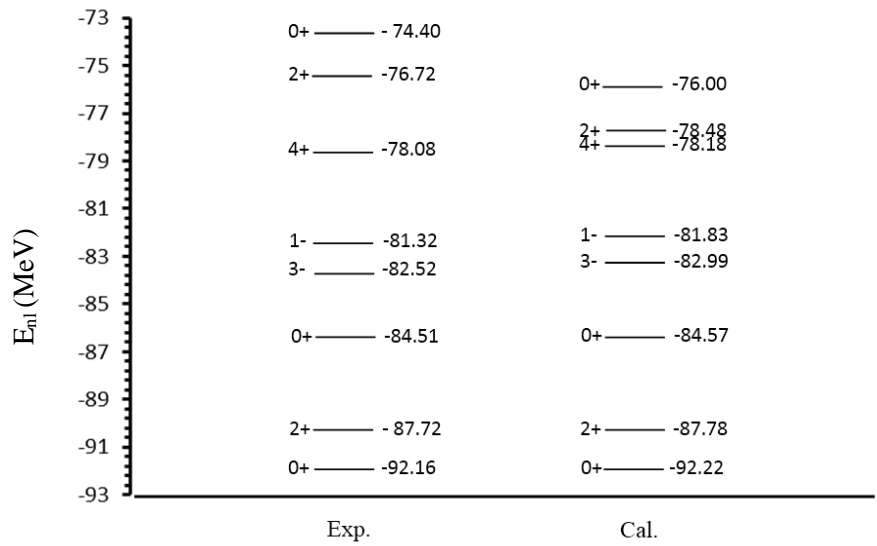

**Fig. 1:** The calculated energy spectrum and compared with the experimental spectrum of $^{12}$C

Using Eq. (12), results of the eigenfunctions of $^{12}$C isotope are found as:

$$\psi_n = N\, x^{-2+\sqrt{4+\xi_3}} e^{-\sqrt{\xi_1} x} L_n^{2\sqrt{4+\xi_3}}(2\sqrt{\xi_1}\, x) \tag{28}$$

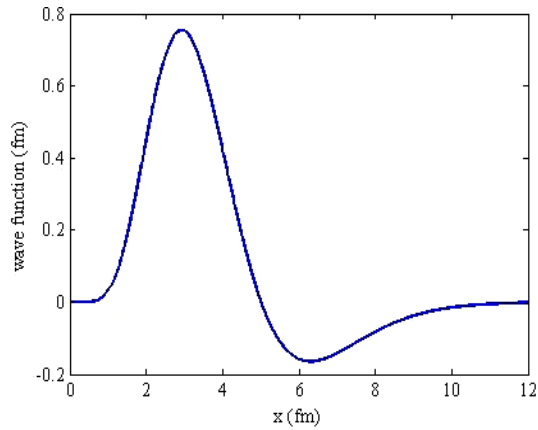

**Fig. 2:** The wave function of the ground state is calculated for $^{12}$C isotope

The wave function of the ground state is drawn at figure 2. It's according to boundary conditions.

## 4. Conclusions

In this paper, through the calculation of the energy levels and the wave functions of $^{12}$C isotope in relativistic cluster model, it was found that the three-alpha cluster model is quite appropriate model to express the structure of $^{12}$C isotope. Also, comparing the experimental data with the results was shown that the modified Yukawa potential plus coulomb potential may be used successfully in the cluster model studies.